# Drift-insensitive distributed calibration of probe microscope scanner in nanometer range: Approach description

Rostislav V. Lapshin[1, 2]

[1]*Solid Nanotechnology Laboratory, Institute of Physical Problems, Zelenograd, Moscow, 124460, Russian Federation*

[2]*Moscow Institute of Electronic Technology, Zelenograd, Moscow, 124498, Russian Federation*

E-mail: rlapshin@gmail.com

A method is described intended for distributed calibration of a probe microscope scanner consisting in a search for a net of local calibration coefficients (LCCs) in the process of automatic measurement of a standard surface, whereby each point of the movement space of the scanner can be defined by a unique set of scale factors. Feature-oriented scanning (FOS) methodology is used to implement the distributed calibration, which permits to exclude *in situ* the negative influence of thermal drift, creep and hysteresis on the obtained results. The sensitivity of LCCs to errors in determination of position coordinates of surface features forming the local calibration structure (LCS) is eliminated by performing multiple repeated measurements followed by building regression surfaces. There are no principle restrictions on the number of repeated LCS measurements. Possessing the calibration database enables correcting in one procedure all the spatial distortions caused by nonlinearity, nonorthogonality and spurious crosstalk couplings of the microscope scanner piezomanipulators. To provide high precision of spatial measurements in nanometer range, the calibration is carried out using natural standards – constants of crystal lattice. The method allows for automatic characterization of crystal surfaces. The method may be used with any scanning probe instrument.

## 1. Introduction

Usually, a probe microscope scanner is characterized by three calibration coefficients $K_x$, $K_y$, $K_z$ representing sensitivities of X, Y, Z piezomanipulators, respectively (to take into consideration a possible nonorthogonality of X, Y piezomanipulators, an obliquity angle should be additionally determined) [1, 2, 3, 4]. Because of piezomanipulator's nonlinearity [5, 6] and spurious crosstalk couplings, the probe microscope scanner may be described by the above coefficients only near the origin of coordinates, where the influence of the distortion factors is insignificant. As moving away from the origin of coordinates, the topography measurement error would noticeably increase reaching the utmost value at the edge of the scanner field [7].

The problem may be solved by using a distributed calibration, which implies determining three local calibration coefficients (LCCs) $K_x$, $K_y$, $K_z$ for each point of the scanner movement space, which can be thought of as scale factors for axes *x*, *y* and *z*, respectively [8, 9, 10, 11]. A reference surface used for calibration should consist of elements, called hereinafter features, such that the distances between them or their sizes are known with a high precision. The corrected coordinate of a point on the distorted image of an unknown surface is obtained by summing up the LCCs related to the points of the movement trajectory of the scanner (see Secs. 2.2, 3).

# Drift-insensitive distributed calibration of probe microscope scanner

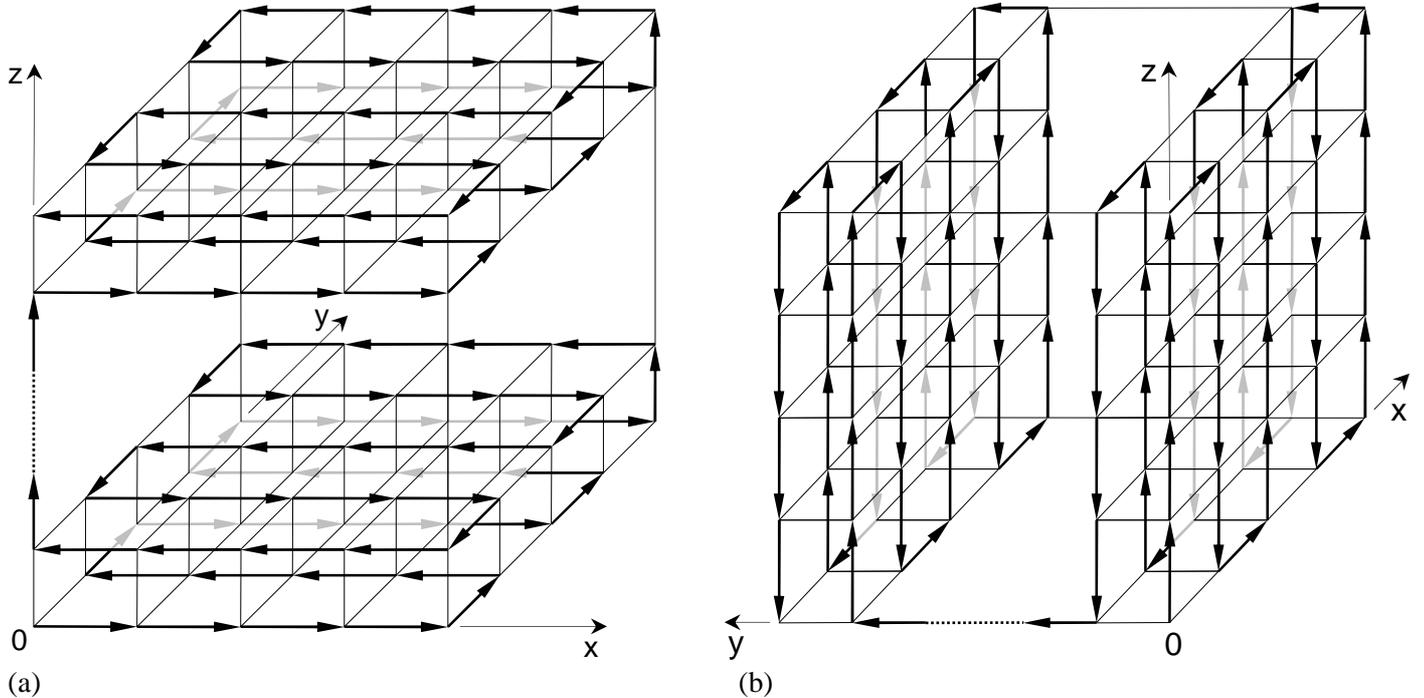

(a)                                                      (b)

Fig. 1. Partitioning of the scanner movement space with a net, LCCs being sought for in the vicinity of each node. The arrows show a spatial movement trajectory by the net nodes during the distributed calibration. (a) Calibration in the lateral planes with "fixed" positions of the scanner Z manipulator. (b) Calibration in the vertical planes with fixed positions of the scanner Y manipulator. In both cases, position of the scanner Z manipulator is set up by means of a coarse approach stage. First, a local scanning of the standard surface is executed in the vicinity of each net node, then the nearest LCS is detected, by which the LCCs are determined. In order to reduce the resulting creep, the movement trajectory from one node to the other is chosen so that the movement in the adjacent (a) lines, (b) columns and in the adjacent (a) horizontal, (b) vertical planes be implemented in the mutually opposite directions.

Both lumped and distributed calibration of the probe microscope scanner should be carried out by the data where distortions caused by drifts (thermal drifts of instrument components plus creeps of piezomanipulators) are eliminated. Otherwise, the measurements will have large errors [2, 3, 12, 13, 14, 15, 16, 17]. In the present work, to eliminate the negative influence of thermal drift and creep on the distributed calibration results, the methods are used of feature-oriented scanning (FOS) and of counter-scanning suggested in Refs. 15, 16 (see Sec. 2).

It is necessary to note that manual calibration is quite a laborious process even in case the scanner is characterized by the three parameters. As to the search for distributed coefficients, it is only possible under full automatic control.

The investigation of the distributed calibration technique as a whole consists of three parts. In the first part presented here, the task of distributed calibration is formulated and its solution based on the FOS-approach is described. Simultaneously with the description of the performed operations, key notions and techniques underlying FOS are shortly introduced and explained. The second part of the investigation is given in Ref. 12, it is devoted to the so-called virtual calibration mode. With this mode, instead of measurement of the real surface of a standard, the calibration program performs "measurement" of an image of the standard surface, which has been obtained earlier during a regular scanning. The virtual mode is intended for simulating the process of calibration and validating the analytical solutions found in the present work. The use of the virtual mode allows for analyzing the operation of a probe microscope scanner, detecting and estimating the errors that occurred. The third part of the investigation is given in Ref. 7, where the experimental results obtained during real distributed scanner calibration by crystal lattice of highly oriented pyrolytic graphite (HOPG) are presented and discussed.

## 2. Measurement process of a standard surface

Shortly, the suggested method of distributed calibration [10] of a probe microscope scanner is as follows. First, the scanning field is "covered" with a square cell net where the nodes correspond to the absolute integer coordinates $x$, $y$, $z$ of the scanner (see Fig. 1). During the calibration, a determination of LCCs is carried out in the neighborhood of each net node. For this purpose, a sequence of the three basic measurement operations is carried out at a standard surface: probe attachment, aperture scanning, and skipping.

The attachment of the microscope probe to a surface feature allows capturing the feature located near the current net node and to holding it within the "field of view" of the instrument. Scanning a small neighborhood called an aperture around the captured feature followed by recognition of topography in the obtained scan enables detecting local calibration structure (LCS). LCS usually consists of three features A, B and C such that the distances between them are *a priori* known precisely. To get the sought-for LCCs, it is required to measure the distances between the features A and B, A and C. Drift-insensitive measurement of these distances is carried out by means of the skipping operation.

### 2.1. Detailed description of the calibration operations

Let us consider the process of the distributed calibration in detail. The microscope probe moves by the net nodes like by a raster (see Fig. 1(a)). The size of the net square side approximately defines how smoothly the LCCs are changing within the found distribution. A pause is inserted after each movement (see Fig. 2, pos. 1), during which the piezoscanner creep produced by this movement is decaying. The larger the distance between the nodes of the initial net is, the longer the pause is set.

#### 2.1.1. Probe attachment

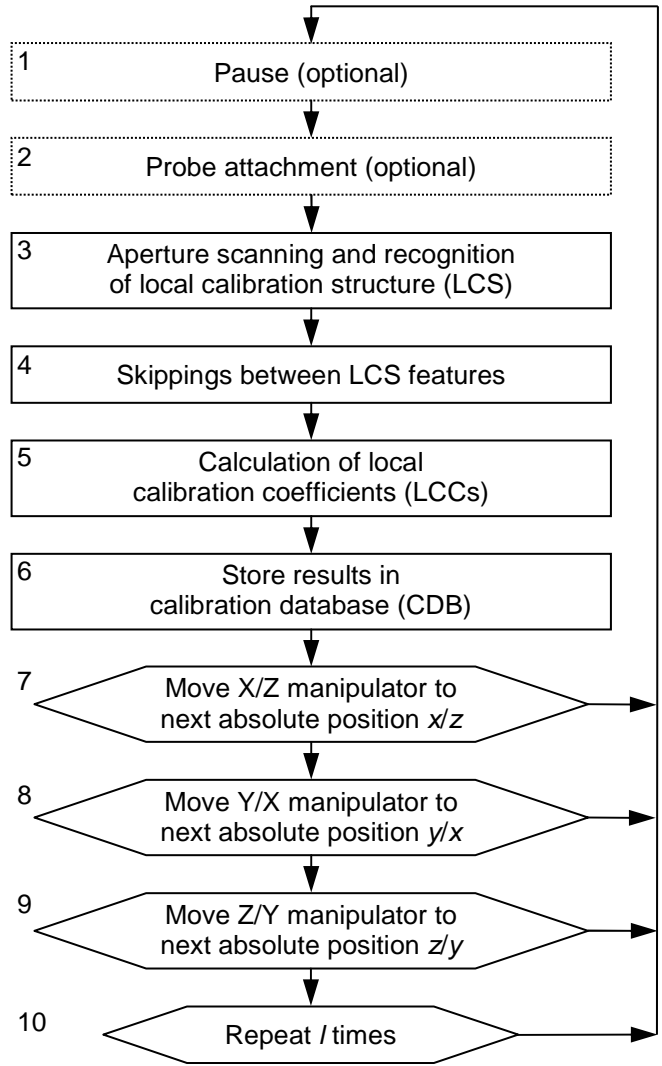

Fig. 2. Simplified flowchart of distributed calibration of the probe microscope scanner. The designations at the left of the slash (see pos. 7-9) correspond to calibration in the lateral planes with "fixed" positions of Z manipulator (see Fig. 1(a)), the ones at the right correspond to calibration in the vertical planes with fixed positions of Y manipulator (see Fig. 1(b)). Absolute position of the scanner Z manipulator is adjusted by means of a coarse approach stage. LCCs are searched for in the vicinity of absolute integer coordinates $x$, $y$, $z$ of the scanner. In some practical cases, pause and probe attachment may be omitted. To increase accuracy, the calibration is repeated $l$ times. The number $l$ of repeated calibrations has no limitations. In the CDB stored are: the absolute real coordinates of points of the scanner space for which the LCCs were obtained and the LCCs values themselves corresponding to these points.

When the pause is over, the feature A of the standard surface nearest to the current net node ($x$, $y$) is being "captured" (see Fig. 3(a)). Topography elements that look like hills or pits may be used as features of the standard surface. Those may be for example: atoms, interstices, molecules, clusters, grains, nanoparticles, crystallites, quantum dots, nanoislets, nanopillars, pores, etc. [15].

**Drift-insensitive distributed calibration of probe microscope scanner**

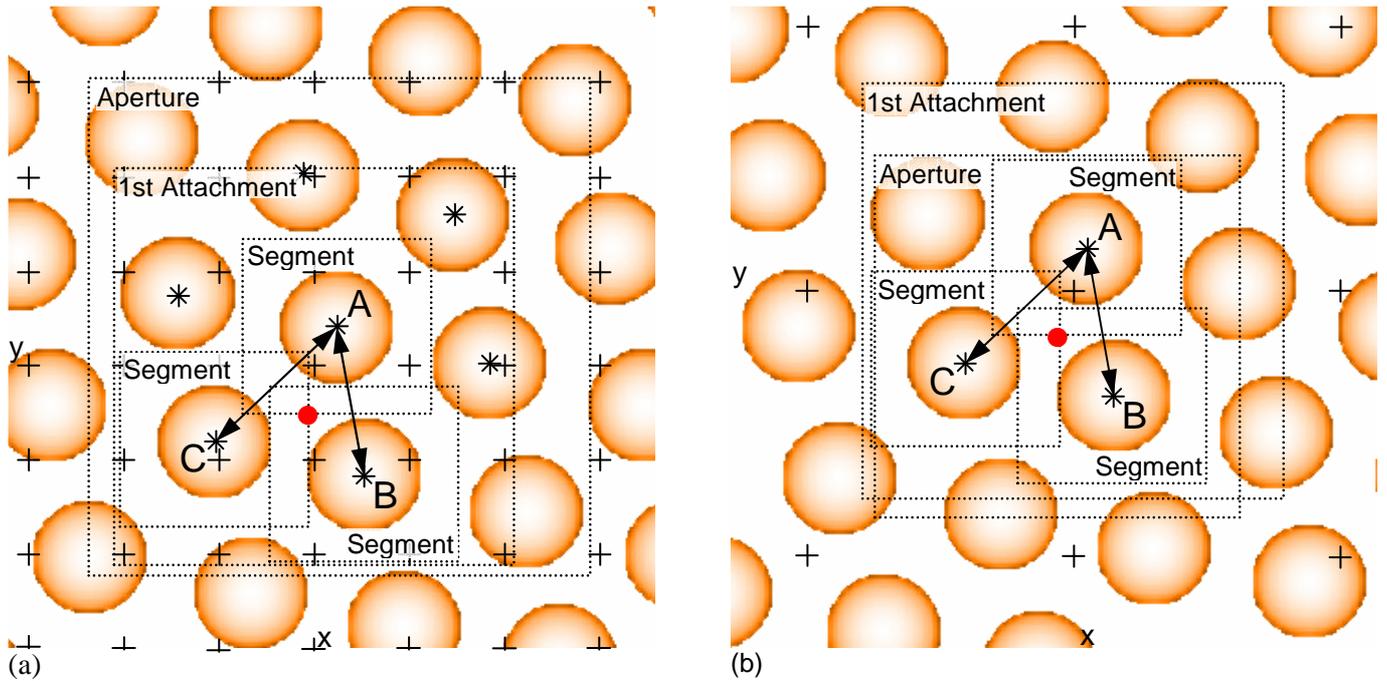

(a) (b)

Fig. 3. Schematic image of the standard surface. Nodes of the net dividing the scanner field into square cells are designated as "+". The nodes correspond to the absolute integer coordinates of the scanner. The point ($x$, $y$) is the current net node, the LCCs are searched within its vicinity. The group of three features A, B and C of the standard surface serves as the LCS, the distances between the features or sizes of the features are known *a priori* with a high accuracy. Positions of the features detected during calibration are designated as "∗". LCCs are determined for the "gravity center" of the LCS ABC designated as "•". The attachment procedure scans a square neighborhood of the current net node in order to detect the nearest (a) feature (designated as A), (b) LCS (designated as A, B, C). The aperture is a square scan containing the captured (a) feature A, (b) LCS ABC in its center. The aperture size is set so as to enclose at least one LCS. The aperture is intended for approximate determination of relative coordinates of the features (a) B and C, (b) A, B and C. The arrows ↔ between the features A and B, A and C symbolize skipping. With skipping, a segment is used – a square scan of the least sizes enclosing just one feature.

The feature capture (tracking) is carried out by scanning a small square neighborhood around the feature, recognizing surface features on the obtained local scan [1, 15] and then moving the probe to the position of the feature located nearest of all to the local scan center. The described sequence of operations is known as probe attachment (see Fig. 2, pos. 2) [15]. The captured feature is then being held for some time within the field of view of the instrument by means of successively repeated attachments.

When the feature is being captured for the first time, a square area is scanned out of such size that being located arbitrary relative to the structure of the standard surface it is able to enclose at least one feature of the standard surface (see Fig. 3(a), "1st Attachment"). To increase the productivity, while executing next attachments, the sizes of the scan area is reduced to the sizes of a segment, which is a square scan enclosing one feature only (see Fig. 3(a)) [15].

During the attachments, the drift velocity is determined [15, 16]. If the drift velocity turns out to be greater by module than a certain preset value then the attachment is repeated. Usually, the preset value is chosen to represent the mean velocity typical of the given microscope with which the microscope drifts after warming up [7, 12]. Thus, the attachment allows to determine the end moment of the creep induced by movement to the next net node. In case of a large net step and corresponding strong creep excitation, it is recommended to set a longer pause in order to prevent substantial deviation of the scanner from the accepted movement trajectory by net nodes (see Fig. 1) while performing series of attachments. The fact is that with a substantial deviation, the following scanner return to this trajectory will excite a strong creep in its turn.

*2.1.2. Aperture scanning and LCS recognition*

At the next step, an aperture [15] scanning is carried out (see Fig. 2, pos. 3). The aperture is a square area around the captured feature A which encloses several neighboring features (see Fig. 3(a)). During the recognition of the obtained aperture, coordinates of the neighboring features are determined (approximately) relative to feature A position. Among the detected neighboring features, features B and C are defined to compose, along with the feature A, the local calibration structure ABC [10, 15].

*2.1.3. Skipping of LCS features*

After the LCS ABC has been detected, skippings (see Fig. 2, pos. 4) [15] between feature A and its neighbors B and C are carried out, one after another (see Fig. 3). The skipping (designated as ↔) is a basic FOS measurement operation intended, in particular, for accurate determination of the relative coordinates of the neighboring features.

The feature skipping operation A↔B (skipping cycle) consists in moving the probe from feature A position to feature B position, scanning-recognizing the segment of feature B, calculating the "forward" differences of coordinates of features A and B, return to position of feature A, scanning-recognizing its segment and then calculating the "backward" differences of coordinates of features A and B. The relative coordinates of features A and B are calculated as a half-sum of the obtained forward and backward differences. Such approach permits to exclude distortions produced by the drift of the microscope probe relative to the sample surface [15].

The skipping provides accurate results under the condition of constant drift velocity [15]. The skipping operation is repeated several times [7], and then the obtained relative coordinates of the features are averaged out. Once the accurate coordinates of the features forming LCS have been defined, the sought LCCs $K_x$, $K_y$, $K_z$ are calculated (see Fig. 2, pos. 5).

Provided that the distance between the net nodes is comparable with the distance at which the scanner is being displaced during the skipping, the pause insertion and the attachment-to-feature operation may be omitted in a number of cases. The fact is that the aperture scanning followed by aperture recognition is actually a sort of attachment. Since the aperture sizes are greater than the sizes of the scanning area typical of a regular attachment, the attachment made with the help of aperture just provides a somewhat less accurate holding of the captured feature in the center of the instrument's field of view.

For the absolute coordinates *x*, *y*, *z* of the LCS to which the LCCs $K_x(x, y, z)$, $K_y(x, y, z)$, $K_z(x, y, z)$ are related, it is suitable to use the "gravity center" coordinates of the calibration structure. If several calibration structures are located near the current net node then, after aperture recognition, such calibration structure is selected among them whose coordinates are the closest of all to the node (see Fig. 3(a)).

*2.2. Using LCCs for determination of the true scanner movements*

Generally, the true (corrected) coordinates $\bar{x}$, $\bar{y}$, $\bar{z}$ of a point *x*, *y*, *z* in the image of an unknown surface are found by summing up the LCCs of the points of the trajectory by which the scanner has moved into the given image point

$$\begin{aligned}
\bar{x}_i(x_i, y_i, z_i) &= \sum_i K_x(x_i, y_i, z_i), \\
\bar{y}_i(x_i, y_i, z_i) &= \sum_i K_y(x_i, y_i, z_i), \\
\bar{z}_i(x_i, y_i, z_i) &= \sum_i K_z(x_i, y_i, z_i),
\end{aligned} \quad (1)$$

where $x_i$, $y_i$, $z_i$ are integer-valued coordinates of *i*th point of the movement trajectory.

**Drift-insensitive distributed calibration of probe microscope scanner**

Actually, while summing up LCCs, the local unit steps corresponding to each trajectory point are summed. LCCs have a plus sign while moving in the positive direction and a minus sign in the negative direction. As in general case the LCCs depend on more than one coordinate, the correction result for a point of the scanner field is, strictly speaking, path-dependent, i. e., it would depend on a spatial path the scanner had passed before getting to this point.

*2.3. Types of movement trajectory by the net nodes*

In the suggested method of distributed calibration, LCCs are determined using the following two types of trajectory of movement by the net nodes:

(1) With "fixed" positions of Z manipulator, coordinates of X and Y manipulators are changed as shown in Fig. 1(a);

(2) With fixed positions of Y manipulator, coordinates of Z and X manipulators are changed as depicted in Fig. 1(b).

In Fig. 2 (pos. 7-9), designations at the left of a slash correspond to the trajectory of the first type, designations at the right – the second type. In both cases, the movements by the net nodes in neighboring lines/columns and in neighboring horizontal/vertical planes are carried out in opposite directions. Moving in opposite directions permits to decrease the resulting creep produced by the movement by the net nodes along directions $x$, $y$ and $z$, $x$, respectively.

It should be noted that the productivity of calibration with fixed Y manipulator is higher in comparison with fixed Z manipulator since $n$ times less aperture scans-recognitions are required for each net node, where $n$ is the number of fixed positions of Z manipulator. However, as it was pointed out above, beginning with a certain moment, a violation of the accepted trajectory of movement by the net nodes occurs because of the long holding of the same feature/LCS.

*2.4. Introducing nonorthogonality and spurious couplings*

By using lateral LCCs $K_x$, $K_y$ of the orthogonal scanner [1], only static nonlinear raster distortions resulted from a nonlinear response of scanner piezoceramics to the applied voltage can be corrected. To also correct residual nonorthogonality and static nonlinear raster distortions resulted from local imperfections of movement straightness of the scanner X, Y manipulators (guidance errors; caused by spurious couplings X⇔Y, Z⇒X, Z⇒Y), the calibration coefficients $\overline{K}_x$, $\overline{K}_y$ of the nonorthogonal scanner and the obliquity angle $\alpha$ (scanner nonorthogonality) [1, 2, 3, 4, 18, 19, 20] calculated by LCSs in many points of the scanner field should be used.

In the simplest case (see Fig. 3), when three features A, B, C with *a priori* precisely known distances between them are used as LCS [1], it is sufficient to carry out two skippings: A⇔B and A⇔C. For example, it applies to the case of three neighboring carbon atoms on surface of HOPG monocrystal [1].

If feature sizes of the standard surface are *a priori* known precisely [19], it is sufficient, instead of scanning the aperture and performing two skippings, to carry out counter-scanning [16] of the segment of a single feature A. When making a counter-scanning, first, a conventional image is obtained called the direct image, after that a counter image is obtained by reversing the direction of the probe movement along a line and the direction of the probe movement from line to line [2, 14, 16]. The found pair of images is called counter-scanned images (CSIs) [16]. It was shown in Ref. 16 that in order to make drift correction of an SPM image possible, it is sufficient to have at least one common surface feature between the direct and the counter images. So, after drift in the segment

**R. V. Lapshin**

has been corrected, the true sizes of feature A may be determined, and then the sought for LCCs corresponding to the absolute position of feature A may be calculated.

Implementation of such method of calibration in the lateral plane requires three different lateral sizes of the feature to be known [1, 4]; calibration in the vertical plane requires one vertical size (for example, step height when using atoms/interstices as features) [19, 21, 22]. The advantage of calibration by known feature sizes consists in a substantially greater productivity. Provided that both the distance between LCS features and the sizes of the features of the standard surface are all known, the precision of distributed calibration may be improved. To do this, instead of a regular scanning during the skipping operation, a counter-scanning of the segments should be performed followed by determining the LCCs by the above two methods.

The distribution of LCCs in the plain field is obtained by performing scanner calibration for some "fixed" position of Z manipulator. Beside nonlinearity and nonorthogonality of the scanner, mutual spurious couplings of X⇔Y type are also taken into account in the found distribution. The change in Z manipulator position while moving in the lateral plane from one LCS to another shows the value of error induced by spurious couplings of X⇒Z, Y⇒Z types (the bowl-shaped image effect well noticeable on flat samples) [10, 19, 20, 23].

The spatial distribution of LCCs $\overline{K}_x$, $\overline{K}_y$ can be obtained by performing scanner calibration for different positions of Z manipulator set by the coarse approach stage. Beside nonlinearity in the vertical plane, the spurious couplings of Z⇒X and Z⇒Y types will be accounted in the found distribution. In general, the spurious couplings X⇔Y, Z⇔X, Z⇔Y arise due to imperfections in construction, control, materials, and manufacturing process of the scanner.

*2.5. Calibration database*

The LCCs obtained during the distributed calibration along with the absolute coordinates of LCS to which they correspond are stored in the calibration database (CDB, see Fig. 2, pos. 6). By storing the dates of the local calibrations in a database, it is possible to accumulate the information on repeated calibrations for several years, which allows determining the scanner piezoceramics age dynamics [7]. As the age dynamics are known, it is easy to find such time interval during which the accumulated LCCs would yield a regression surface (see Sec. 3) ensuring correction of the surface image with some minimum error.

*2.6. Adaptive properties of the algorithm*

The modulus of drift velocity during attachment, skipping or counter-scanning should not exceed a certain value – the one at which the current feature can go out of the limits of the scanned segment. Another condition imposed on the drift is invariability of the drift velocity during one cycle of skipping or segment counter-scanning [15, 16]. The better the last condition is satisfied, the less error of drift elimination may be obtained which, in its turn, would lead to a more accurate LCC determination. Measurements show that the drift velocity remains practically constant over tens of seconds [14, 15, 16] or even tens of minutes [24] whereas one cycle of skipping between neighboring carbon atoms on pyrolytic graphite surface may take as little as 300 ms [15].

To eliminate the calibration errors connected with large changes in drift velocity, a ceaseless *in situ* monitoring of drift velocity is established while skipping and/or counter-scanning [15, 16]. Once an unallowable change in drift velocity is registered, the data obtained during the current skipping cycle are declared spoiled and discarded. After that, several idle skipping cycles are inserted. If the drift velocity has recovered during that time, the inter-

**Drift-insensitive distributed calibration of probe microscope scanner**

rupted skipping is continued, otherwise it is restarted. In case of a long-term instability, the microscope probe is returned to the current net position and the local calibration is carried out once again.

If an unallowable change of the drift velocity interrupts the calibration by feature sizes, then an idle counter-scanning of the segment is applied. Here, like with the skipping, the local calibration for the current net node will be restarted if the drift velocity is unable to stabilize during the set time interval. The suggested calibration algorithm can thus be considered as adaptive, as it changes its operation in accordance with the actual changes in measurement conditions.

## 3. Correction of scan of an arbitrary surface

Because of the drift as well as the differences in dimension, structure, and orientation between the cells of the initial net and the lattice of the standard surface, the obtained LCC net is no longer a square cell net with integer period, and so it should be transformed to one for subsequent use. The transformation consists in constructing a regression surface, i. e., drawing through the noisy LCCs such a smooth surface $f$ that a sum of squared deviations between the LCC values and the corresponding surface points is minimal. These constructions may be carried out, for instance, for each fixed position of the scanner Z manipulator. As a result, the following set of surfaces is obtained

$$\begin{aligned}\overline{K}_x^r &= f_{\overline{K}_x}(x, y, z_i), \\ \overline{K}_y^r &= f_{\overline{K}_y}(x, y, z_i), \\ \alpha^r &= f_\alpha(x, y, z_i), \\ K_z^r &= f_{K_z}(x, y, z_i),\end{aligned} \quad (2)$$

where letter $r$ means regression and $z_i$ are fixed positions of the scanner Z manipulator ($i=1, 2, …, n$) adjusted during calibration by means of coarse approach stage displacement. Constructing regression surfaces (2) also permits to reduce the influence of errors of determination of LCCs and local obliquity angles [1] on the results of the nonlinear correction.

Formally, in any $(x, y)$ point the pair of coefficients $K_x$, $K_y$ of some hypothetical orthogonal scanner may be expressed in terms of the calibration coefficients $\overline{K}_x$, $\overline{K}_y$, and the obliquity angle $\alpha$ by using the following transformation that establishes a relationship between the unit length scales of a rectangular and an oblique system of coordinates [1]

$$\begin{aligned}K_x(x, y) &= \overline{K}_x(x, y) + \overline{K}_y(x, y)\sin[\alpha(x, y)], \\ K_y(x, y) &= \overline{K}_y(x, y)\cos[\alpha(x, y)].\end{aligned} \quad (3)$$

The $z$ coordinate in transformations (3) is omitted to simplify the notation.

By replacing LCCs $\overline{K}_x$, $\overline{K}_y$, and obliquity angle $\alpha$ in formulae (3) with the values taken from the corresponding regression surfaces (2) and then substituting the obtained expressions for $K_x$, $K_y$ in (1), one may write

$$\begin{aligned}\overline{x}_i(x_i, y_i) &= \sum_i \{\overline{K}_x^r(x_i, y_i) + \overline{K}_y^r(x_i, y_i)\sin[\alpha^r(x_i, y_i)]\}, \\ \overline{y}_i(x_i, y_i) &= \sum_i \overline{K}_y^r(x_i, y_i)\cos[\alpha^r(x_i, y_i)],\end{aligned} \quad (4)$$

where $x_i$, $y_i$ are integer-valued lateral coordinates of the $i$th point of the movement trajectory.

Strictly speaking, the $\overline{K}_x^r(x_i, y_i)$ term in transformations (4) is responsible for a shift of the point to be cor-

**R. V. Lapshin**

rected along *x* if only the movement trajectory of the scanner has an *x*-component at this point. The $\overline{K}_y^r(x_i, y_i)\sin[\alpha^r(x_i, y_i)]$ term sets a shift of the point to be corrected along *x* if only the movement trajectory of the scanner has an *y*-component at this point. The $\overline{K}_y^r(x_i, y_i)\cos[\alpha^r(x_i, y_i)]$ term provides a shift of the point under correction along *y* if only the movement trajectory of the scanner has an *y*-component at this point. Actually, the above conditions establish the rules for calculating the corrected coordinates $\bar{x}$, $\bar{y}$. To represent these rules analytically, formulae (4) should be rewritten as follows

$$\bar{x}_i(x_i, y_i) = \sum_i \{\overline{K}_x^r(x_i, y_i)\Delta x_i + \overline{K}_y^r(x_i, y_i)\sin[\alpha^r(x_i, y_i)]\Delta y_i\}$$
$$\bar{y}_i(x_i, y_i) = \sum_i \overline{K}_y^r(x_i, y_i)\cos[\alpha^r(x_i, y_i)]\Delta y_i,$$
(5)

where $\Delta x_i = x_i - x_{i-1}$, $\Delta y_i = y_i - y_{i-1}$ are differences of integer-valued coordinates of the trajectory neighboring points, which may take the values 0 or ±1.

It might seem that we could use one regression surface less if we calculate LCCs $K_x(x, y)$, $K_y(x, y)$ according to formula (3), replace them with corresponding regression surfaces $K_x^r(x, y)$, $K_y^r(x, y)$, and then substitute in (1). However, we must not do this. The fact is that the "switches" $\Delta x_i$, $\Delta y_i$ provide the information of the trajectory, which would be fixed in this case. So the CDB will be bound to a single particular trajectory and, as a result, will lose its universality and therefore its practical value. Thus, LCCs $\overline{K}_x$, $\overline{K}_y$, and local obliquity angle $\alpha$ should be stored in the CDB and processed separately.

In formulae (5), polynomials are suitable for use as regression surfaces

$$\overline{K}_x^r(x, y) = a_0 + a_1 y + a_2 x + a_3 xy + a_4 y^2 + a_5 x^2 + \ldots = \sum \mathbf{a} x^{\mathbf{p}_{ax}} y^{\mathbf{p}_{ay}},$$
$$\overline{K}_y^r(x, y) = b_0 + b_1 y + b_2 x + b_3 xy + b_4 y^2 + b_5 x^2 + \ldots = \sum \mathbf{b} x^{\mathbf{p}_{bx}} y^{\mathbf{p}_{by}},$$
(6)
$$\alpha^r(x, y) = c_0 + c_1 y + c_2 x + c_3 xy + c_4 y^2 + c_5 x^2 + \ldots = \sum \mathbf{c} x^{\mathbf{p}_{cx}} y^{\mathbf{p}_{cy}},$$
$$K_z^r(x, y) = d_0 + d_1 y + d_2 x + d_3 xy + d_4 y^2 + d_5 x^2 + \ldots = \sum \mathbf{d} x^{\mathbf{p}_{dx}} y^{\mathbf{p}_{dy}},$$

where $\mathbf{a}=(a_0, a_1, a_2, \ldots)$, $\mathbf{b}=(b_0, b_1, b_2, \ldots)$, $\mathbf{c}=(c_0, c_1, c_2, \ldots)$, $\mathbf{d}=(d_0, d_1, d_2, \ldots)$ are vectors of the polynomial coefficients; $\mathbf{p}_x=(0, 0, 1, 1, 0, 2, \ldots)$, $\mathbf{p}_y=(0, 1, 0, 1, 2, 0, \ldots)$ are vectors of powers of *x*, *y* variables, respectively. The highest degree of the polynomial for each regression surface (6) may be chosen so as to ensure, for example, the least error of the mean value of the calibration size on the corrected image of a standard [12].

Transformations (5) are universal and applicable for coordinate correction of any kind of trajectory. When the trajectory is a raster, transformations (5) take the following form

$$\bar{x}_i(x_i, y_i) = \sum_i \{\overline{K}_x^r(x_0 + x_i, y_0 + y_i)\Delta x_i + \overline{K}_y^r(x_0, y_0 + y_i)\sin[\alpha^r(x_0, y_0 + y_i)]\Delta y_i\}$$
$$\bar{y}_i(y_i) = \sum_i \overline{K}_y^r(x_0, y_0 + y_i)\cos[\alpha^r(x_0, y_0 + y_i)]\Delta y_i,$$
(7)

where $x_0$, $y_0$ are coordinates setting a raster position in CDB relative to the scanner origin of coordinates; $x_i$, $y_i$ are points of the raster trajectory of the distorted image which belong to the ranges 0, 1, 2, …, $x_{max}$ and 0, 1, 2, …, $y_{max}$, respectively ($x_{max}+1$ is the number of points in the raster line and $y_{max}+1$ is the number of raster lines).

Formulae (7) are intended for correction of raster trajectory of the direct image (designated as number 1 in Fig. 1(a) in [16]). Proceeding by analogy, it is easy to obtain formulae that correct the raster trajectory of the

**Drift-insensitive distributed calibration of probe microscope scanner**

counter image (designated as number 2 in Fig. 1(a) in [16]). To do that, one should consider the opposite direction of the trajectory and its origin location in the coincidence point ($x_{max}$, $y_{max}$) [16].

With raster scanning, the trajectory pattern is known *a priory* and may be easily formalized. As retrace is used in a line, it allows passing from trajectory points $x_i$, $y_i$ to arbitrary raster points $x$, $y$, the formulae (7) take the form

$$\bar{x}(x,y) = \sum_x \bar{K}_x^r(x_0+x, y_0+y) + \sum_y \bar{K}_y^r(x_0, y_0+y)\sin[\alpha^r(x_0, y_0+y)],$$

$$\bar{y}(y) = \sum_y \bar{K}_y^r(x_0, y_0+y)\cos[\alpha^r(x_0, y_0+y)].$$
(8)

Here, $\Delta x$, $\Delta y$ differences are present in formulae (8) implicitly through the known movement trajectory in the raster.

Instead of sum accumulation in expressions (8), the sought coordinates may be determined directly by calculating the following integrals

$$\bar{x}(x,y) = \int_0^x \bar{K}_x^r(x_0+x, y_0+y)dx + \int_0^y \bar{K}_y^r(x_0, y_0+y)\sin[\alpha^r(x_0, y_0+y)]dy,$$

$$\bar{y}(y) = \int_0^y \bar{K}_y^r(x_0, y_0+y)\cos[\alpha^r(x_0, y_0+y)]dy.$$
(9)

The formulae for counter image are obtained likewise.

For regression surfaces represented by polynomials (6), the integrals (9) have analytical solutions. For planes, for example, the following formulae can be obtained

$$\bar{x}(x,y) = [a_0 + a_1(y_0+y) + a_2 x_0]x + \frac{a_2}{2}x^2$$

$$+ \frac{1}{c_1}\left[ 2\sin\left(\frac{c_1 y}{2}\right) \left\{ \begin{array}{l} (b_0 + b_1 y_0 + b_2 x_0)\sin\left[c_0 + c_1\left(y_0+\frac{y}{2}\right) + c_2 x_0\right] \\ + \frac{b_1}{c_1}\cos\left[c_0 + c_1\left(y_0+\frac{y}{2}\right) + c_2 x_0\right] \end{array} \right\} \right.$$
$$\left. - b_1 y \cos[c_0 + c_1(y_0+y) + c_2 x_0] \right],$$
(10)

$$\bar{y}(y) = \frac{1}{c_1}\left[ 2\sin\left(\frac{c_1 y}{2}\right) \left\{ \begin{array}{l} (b_0 + b_1 y_0 + b_2 x_0)\cos\left[c_0 + c_1\left(y_0+\frac{y}{2}\right) + c_2 x_0\right] \\ - \frac{b_1}{c_1}\sin\left[c_0 + c_1\left(y_0+\frac{y}{2}\right) + c_2 x_0\right] \end{array} \right\} \right.$$
$$\left. + b_1 y \sin[c_0 + c_1(y_0+y) + c_2 x_0] \right].$$

For higher order polynomials, the analytical solutions become yet more cumbersome. To avoid intricate formulae, the integrals can be calculated by applying calculus of approximations. Moreover, by using in the integrals (9) instead of $\bar{K}_y^r(x,y)\sin[\alpha^r(x,y)]$ and $\bar{K}_y^r(x,y)\cos[\alpha^r(x,y)]$ the corresponding regression surfaces

$$\bar{K}_{yx}^r(x,y) = \sum \mathbf{e} x^{p_{ex}} y^{p_{ey}},$$
$$\bar{K}_{yy}^r(x,y) = \sum \mathbf{f} x^{p_{fx}} y^{p_{fy}}$$
(11)

(vectors **e**, **f** and **p** are defined similar to those in (6)) built respectively for products $\bar{K}_y(x,y)\sin[\alpha(x,y)]$ and $\bar{K}_y(x,y)\cos[\alpha(x,y)]$ (the analogous substitution may be used in formulae (7), (8)), a more compact formulae may be written down

**R. V. Lapshin**

$$\bar{x}(x,y) = \sum \frac{\mathbf{a}}{\mathbf{p_{ax}}+1} \left[(x_0+x)^{\mathbf{p_{ax}}+1} - x_0^{\mathbf{p_{ax}}+1}\right](y_0+y)^{\mathbf{p_{ay}}} + \sum \frac{\mathbf{e}}{\mathbf{p_{ey}}+1} x_0^{\mathbf{p_{ex}}} \left[(y_0+y)^{\mathbf{p_{ey}}+1} - y_0^{\mathbf{p_{ey}}+1}\right]$$

$$\bar{y}(y) = \sum \frac{\mathbf{f}}{\mathbf{p_{fy}}+1} x_0^{\mathbf{p_{fx}}} \left[(y_0+y)^{\mathbf{p_{fy}}+1} - y_0^{\mathbf{p_{fy}}+1}\right] \quad (12)$$

Strictly speaking, the numerical values obtained by formulae (12) are not equivalent to the values obtained by formulae (9) but are pretty close to them.

## 4. Estimation of scanner errors

The more symmetrical scanner construction is used, and the more precise its fabrication, more homogeneous material, and more evenly distributed load, the more symmetrical the surfaces (2) are and the stronger they resemble solids of revolution. Asymmetry or deviation from a revolution solid may serve as signs of constructional, technological or material imperfection of the scanner or incorrect scanner control. In absence of nonlinearities and spurious crosstalk couplings, the regression surfaces (2) degenerate into horizontal planes. Thus, the difference between the obtained regression surface and the horizontal plane may serve as a measure of the nonlinear distortions and spurious couplings of the scanner [7, 12]. To get estimate for this measure, the following maximum differences will be calculated

$$\Delta^{\max}_{\bar{K}^r_x} = \max \left|\bar{K}^r_x(x,y) - \bar{K}^r_x(0,0)\right|,$$
$$\Delta^{\max}_{\bar{K}^r_y} = \max \left|\bar{K}^r_y(x,y) - \bar{K}^r_y(0,0)\right|, \quad (13)$$
$$\Delta^{\max}_{\alpha^r} = \max \left|\alpha^r(x,y) - \alpha^r(0,0)\right|,$$

where point (0, 0) corresponds to the origin of coordinates of the direct/counter image in case of the virtual calibration [12] or to the scanner's origin of coordinates in case of the real calibration [7].

To estimate root-mean-square deviations of LCCs and obliquity angle the following formulae will be used

$$\sigma_{\bar{K}_x} = \sqrt{\frac{1}{n-1} \sum_{i=1}^{n} \left[\bar{K}_x(x_i,y_i)_i - \bar{K}^r_x(x_i,y_i)\right]^2},$$
$$\sigma_{\bar{K}_y} = \sqrt{\frac{1}{n-1} \sum_{i=1}^{n} \left[\bar{K}_y(x_i,y_i)_i - \bar{K}^r_y(x_i,y_i)\right]^2}, \quad (14)$$
$$\sigma_\alpha = \sqrt{\frac{1}{n-1} \sum_{i=1}^{n} \left[\alpha(x_i,y_i)_i - \alpha^r(x_i,y_i)\right]^2},$$

where $x_i$, $y_i$, $\bar{K}_x(x_i,y_i)_i$, $\bar{K}_y(x_i,y_i)_i$, $\alpha(x_i,y_i)_i$ are the coordinates, LCCs, and the local obliquity angle of the $i$th LCS, respectively, taken from the CDB; $n$ is the number of LCSs in CDB.

## 5. Application of feature-oriented positioning technique

In a number of practical cases it is quite enough to calibrate precisely just a small part of the scanner space located in the vicinity of origin of coordinates, where the residual errors are minimal [7]. To perform the required measurements, the area of interest on the surface can be simply transferred to this part of the scanner space by applying feature-oriented positioning (FOP) methods [15].

A successful solution of the problem of finding LCCs distributed in movement space of the scanner mostly depends on the properties of the standard surface used. The latter should possess an invariable structure in each point of the scanner field. In practice, however, defects and residual stresses distort the surface structure of the standard

**Drift-insensitive distributed calibration of probe microscope scanner**

somewhat complicating the distributed calibration on large fields. In this case, a preliminarily selected "perfect" region of the standard surface has to be moved within the scanner space by using the same FOP methods in order to calibrate the whole space only by this small area of the standard surface [15].

In both cases to perform the transport operations mentioned above, a coarse X, Y, Z positioner is required capable of carrying out a long-distance travel with the step comparable to the typical lateral size of the used reference feature [15]. It is important to note that no special requirements of movement precision are to be met by the coarse positioner, since with FOP the entire movement precision is provided by the scanner (fine positioner) and its calibration [15].

## 6. Discussion

Yet another variation of distributed calibration is shown in Fig. 3(b). Here, during the probe attachment or aperture scanning, the whole LCS is involved rather than a separate feature. To be precise, among the features recognized in a local scan, those are being searched for which make up the LCS nearest to the current net node; after that an attachment of the local scan is carried out to the gravity center of the found LCS. This variation of the distributed calibration has a higher productivity since it uses a reduced-size aperture. The method is applicable when the step of the initial net is greater than a typical size of LCS.

In case the step of the net is comparable with the typical LCS size (see Fig. 3(a)), it is possible after skippings A↔B and A↔C to carry out additional skippings between feature A and all the other features detected in the aperture and forming an LCS with feature A. The obtained LCCs should be associated with the coordinates of the gravity centers of the corresponding LCSs. For instance, according to the configuration shown in Fig. 3(a), 6 times less number of movements between the net nodes, waitings, attachments and aperture scannings may be executed during that kind of calibration.

The skipping operation may be omitted at all when no high precision calibration of the scanner is required yet a sizeable scanner field is to be calibrated or when a less accurate calibration is to be performed for a shorter period of time. By applying counter-scanning in the aperture [16] instead of skipping, it is possible to correct drift in the aperture and then determine the sought for LCCs.

With the calibration method suggested (see Fig. 2), the number of skipping cycles (pos. 4, see also Ref. 7) are not recommended to be set large. Otherwise, the microscope probe may shift at a large distance off the current net node due to drift that after moving the probe to the next net node will induce a strong creep resulting in a long pause. Here, the trajectory of movement by the net nodes accepted in this method (see Fig. 1) is violated which generally leads to undesirable "runaway" of the scanner's piezomanipulators. Moreover, with the large number of skipping cycles, the relative distances between the features measured in a sequence of repeated cycles will correspond, owing to drift, to different absolute positions of the scanner, where, strictly speaking, different nonlinear static distortions occur.

Thus, to obtain more accurate values of the noise-sensitive calibration coefficients $\overline{K}_x$, $\overline{K}_y$ and the obliquity angle $\alpha$ [1], the calibration routine should be repeated many times (see Fig. 2, pos. 10) adding each time a new set of LCCs to the database. The same is true with respect to a number of consequently executed counter scans obtained calibrating by sizes of feature of the standard surface.

Similar arguments exist against the attempts to measure LCS with redundancy by performing an additional skipping B↔C followed by least-squares data adjustment as it is accepted in geodesy [25], for example. Neverthe-

R. V. Lapshin

less, for a small drift, the use of this trick allows to increase the precision of measurement of feature coordinates.

## 7. Conclusion

The developed calibration method employs a number of principles and tricks embeded in the FOS methodology [15], namely: handling separate surface features, movement by short distances from one feature to another located nearby, measurement of the relative distances, localization of measurements, multiple repetitions of the measurements, ceaseless probe attachments to the surface features, continuous monitoring of the drift velocity, drift distortion neutralization by means of hierarchically-organized counter movements. Thus, the suggested calibration method can be thought of as another component of the FOS methodology.

## Acknowledgments

This work was supported by the Russian Foundation for Basic Research (project 15-08-00001) and by the Ministry of Education and Science of Russian Federation (contracts 14.429.11.0002, 14.578.21.0059). I thank O. E. Lyapin and Assoc. Prof. S. Y. Vasiliev for their critical reading of the manuscript; Dr. A. L. Gudkov, Prof. E. A. Ilyichev, Assoc. Prof. E. A. Fetisov, and late Prof. E. A. Poltoratsky for their support and stimulation.

**Drift-insensitive distributed calibration of probe microscope scanner**

R. V. Lapshin